\documentclass[table,xcdraw,multirow]{aastex62}

\submitjournal{ApJ}

\shorttitle{Effect of O$_{2}$ on Haze}
\shortauthors{H\"orst et al.}
\begin{document}

\title{Exploring the Atmosphere of Neoproterozoic Earth: The Effect of O$_{2}$ on Haze Formation and Composition}

\correspondingauthor{Sarah H\"orst}
\email{sarah.horst@jhu.edu}

\author[0000-0003-4596-0702]{Sarah M. H\"orst}
\affil{Department of Earth and Planetary Sciences \\
Johns Hopkins University \\
Baltimore, MD, USA}
\affiliation{Cooperative Institute for Research in Environmental Sciences\\
University of Colorado\\
 Boulder, CO, USA}

\author{Chao He}
\affiliation{Department of Earth and Planetary Sciences \\
Johns Hopkins University \\
Baltimore, MD, USA}

\author{Melissa S. Ugelow}
\affiliation{Department of Chemistry and Biochemistry\\
University of Colorado\\
Boulder, CO, USA}

\author{A. Mark Jellinek}
\affiliation{Department of Earth, Ocean and Atmospheric Sciences\\
University of British Columbia\\
Vancouver, BC, Canada.}

\author{Raymond T. Pierrehumbert}
\affiliation{Department of Physics\\
University of Oxford\\
Oxford, UK.}

\author{Margaret A. Tolbert}
\affiliation{Cooperative Institute for Research in Environmental Sciences\\
University of Colorado\\
 Boulder, CO, USA}
\affiliation{Department of Chemistry and Biochemistry\\
University of Colorado\\
Boulder, CO, USA}



\begin{abstract}

Previous studies of haze formation in the atmosphere of the Early Earth have focused on N$_{2}$/CO$_{2}$/CH$_{4}$ atmospheres. Here, we experimentally investigate the effect of O$_{2}$ on the formation and composition of aerosols to improve our understanding of haze formation on the Neoproterozoic Earth. We obtained in situ size, particle density, and composition measurements of aerosol particles produced from N$_{2}$/CO$_{2}$/CH$_{4}$/O$_{2}$ gas mixtures subjected to FUV radiation (115-400 nm) for a range of initial CO$_{2}$/CH$_{4}$/O$_{2}$ mixing ratios (O$_{2}$ ranging from 2 ppm to 0.2\%). At the lowest O$_{2}$ concentration (2 ppm), the addition increased particle production for all but one gas mixture. At higher oxygen concentrations (20 ppm and greater) particles are still produced, but the addition of O$_{2}$ decreases the production rate. Both the particle size and number density decrease with increasing O$_{2}$, indicating that O$_{2}$ affects particle nucleation and growth. The particle density increases with increasing O$_{2}$. The addition of CO$_{2}$ and O$_{2}$ not only increases the amount of oxygen in the aerosol, but it also increases the degree of nitrogen incorporation. In particular, the addition of O$_{2}$ results in the formation of nitrate bearing molecules. The fact that the presence of oxygen bearing molecules increases the efficiency of nitrogen fixation has implications for the role of haze as a source of molecules required for the origin and evolution of life. The composition changes also likely affect the absorption and scattering behavior of these particles but optical properties measurements are required to fully understand the implications for the effect on the planetary radiative energy balance and climate.

\end{abstract}



\section{Introduction} \label{sec:intro}

The possible presence of an organic haze layer in the atmosphere of Archean Earth, similar to the one currently present in the atmosphere of Titan, has been suggested by a number of studies (see e.g., \citet{Sagan:1997, Pavlov:2001, Pavlov:2001b, Haqq:2008, Domagal:2008, Wolf:2010}) and has been of particular interest as a factor in the resolution of the Faint Young Sun Paradox; although the effect of such a haze layer on climate would depend strongly on the altitude of the layer, the size, composition, and morphology of the particles, and the atmospheric gas composition as a haze layer could also cause an anti-greenhouse effect as it does on Titan (see e.g., \citet{Mckay:1999, Haqq:2008, Arney:2016}).  It has also been used to explain measurements of sulfur isotopes (e.g., \citet{Domagal:2008}). The climate of the Proterozoic presents a number of enigmatic features, including major glaciations at the beginning and end and a long period of apparent stasis in between \citep{Pierrehumbert:2011}. Haze particles absorb and scatter light differently than gases and can therefore impact the temperature structure of an atmosphere. Atmospheric hazes can also serve as cloud condensation nuclei (CCN) and therefore can affect cloud formation (including size, altitude, and lifetime) and atmospheric albedo. Organic compounds present in atmospheric hazes are a source of material that may be useful for the origin and/or evolution of life (see e.g., \citet{Horst:2012}). To understand the role atmospheres play in the habitability of a world and the origin of life, we must therefore understand the role of atmospheric hazes. 

Atmosphere simulation experiments designed to investigate haze formation in the atmosphere of the early Earth have focused on an atmosphere that lacked molecular oxygen (O$_{2}$) (see e.g., \citet{Trainer:2004b, NnaMvondo:2005, Trainer:2006, dewitt:2009, Hasenkopf:2010, Fleury:2015, Gavilan:2017, Fleury:2017}). \citet{Trainer:2013earth} provides an excellent review of our understanding of the atmosphere of the early Earth and previous early Earth experiments. Here we present the results of a set of experiments that investigate the effect of O$_{2}$ on haze formation, particle size, and particle composition in CH$_{4}$/CO$_{2}$/O$_{2}$/N$_{2}$ atmospheres to understand haze formation during the Rise of Oxygen on the early Earth, the effect of abiotic O$_{2}$ on haze formation, and to begin thinking more broadly about the implications for the presence of a haze layer in the atmospheres of exoplanets where O$_{2}$ may be present. 

\section{Materials and Experimental Methods}

\subsection{Experimental phase space}

A characteristic feature of present day deep ocean hydrothermal vent plumes hosted in ultramafic rocks within the very slowly spreading parts of the mid-Atlantic ridge is a very large flux of abiotic CH$_{4}$ (e.g., \citet{Charlou:1998}). From experimental studies of olivine serpentinization \citep{Janecky:1986, Berndt:1996} the hydrothermal circulation of seawater with dissolved CO$_{2}$ leads to the conversion of Fe(II) in olivine to Fe(III) in magnetite, and to the production of H$_{2}$ and the release of the observed CH$_{4}$ gas
through ``Fischer-Tropsch'' reactions of the form
\begin{equation}
{\rm (Fe,~Mg)_2SiO_4 + nH_2O +CO_2 = Mg_3Si_2O_5(OH)_4+Fe_3O_4+CH_4}.
\end{equation}
If the resulting flux of abiotic CH$_{4}$ into the deep ocean drives, in turn, a sufficiently large flux of CH$_{4}$ into the atmosphere this process can potentially lead to the production of organic hazes (see e.g., \citet{Haqq:2008}).

The likelihood that such haze production will occur depends on the extent to which mantle rocks at mid-ocean ridges are involved in hydrothermal circulation and, in particular, on the residence time of CH$_{4}$ released into the deep ocean. In the oxygen-rich and biologically productive present day ocean, little if any of this CH$_{4}$ is expected to escape to the atmosphere as the CH$_{4}$ that is not consumed by methanotrophs will be largely oxidized to CO$_{2}$. Furthermore, any remaining CH$_{4}$ that enters the atmosphere will be oxidized within only a few years to a decade. By contrast, a much larger fraction of this CH$_{4}$ is likely to exit the relatively low biomass Archean and Proterozoic oceans and enter into very low oxygen atmospheres with which they are in equilibrium \citep{Lyons:2014}. 

Methane released to the atmosphere before the Great Oxidation Event (2.4-2.1 Ga) (O$_{2}$ $< 10^{-6}$ PAL (present atmospheric level)) as well as during most of the Proterozoic ($10^{-3}<$ O$_{2}$ $< 10^{-2}$ PAL) will have a residence time exceeding $10^3$ years \citep{Haqq:2008}. Constraining the magnitude of the CH$_{4}$ flux to the atmosphere even asymptotically is challenging. In particular, this flux will depend not only on the average rate of abiotic methane production but also on complex factors including the extent to which the ocean is fully oxygenated, on the full length and spreading rate of mid-ocean ridge compared to present day, and the heights to which hydrothermal vent plumes penetrate stabilizing ocean density stratification, as well as their advection by ocean currents (e.g., \citet{Carazzo:2013}).

To explore the potential climate consequences of supercontinental cycles in low O$_{2}$ worlds (such as neoproterozoic Earth and some extrasolar planets), we are interested in understanding how the delivery of this abiotic methane affects haze formation. We consequently chose initial gas concentrations to explore of possible CH$_{4}$/CO$_{2}$ ratios that result from the complex interplay between the mantle, crust, ocean, and atmosphere for O$_{2}$ concentrations relevant to Archean and Proterozoic Earths \citep{Jellinek:2009, Lenardic:2011, Jellinek:2012}.  A broad range of O$_{2}$ concentrations were explored to understand the effect of O$_{2}$ on particle formation and composition, which has not been previously investigated. 

We note here that there are possible biotic and abiotic sources for CH$_{4}$, CO$_{2}$, and even O$_{2}$, which can be produced at low concentrations by photochemistry (see e.g., \citet{Kasting:2003, Wordsworth:2014}). It is for that reason we investigated a range of CH$_{4}$, CO$_{2}$, and O$_{2}$ concentrations to better understand their role in haze formation.

\subsection{Haze Production setup} %done

\begin{figure}
\centering
\resizebox{5.5in}{!}
{\includegraphics{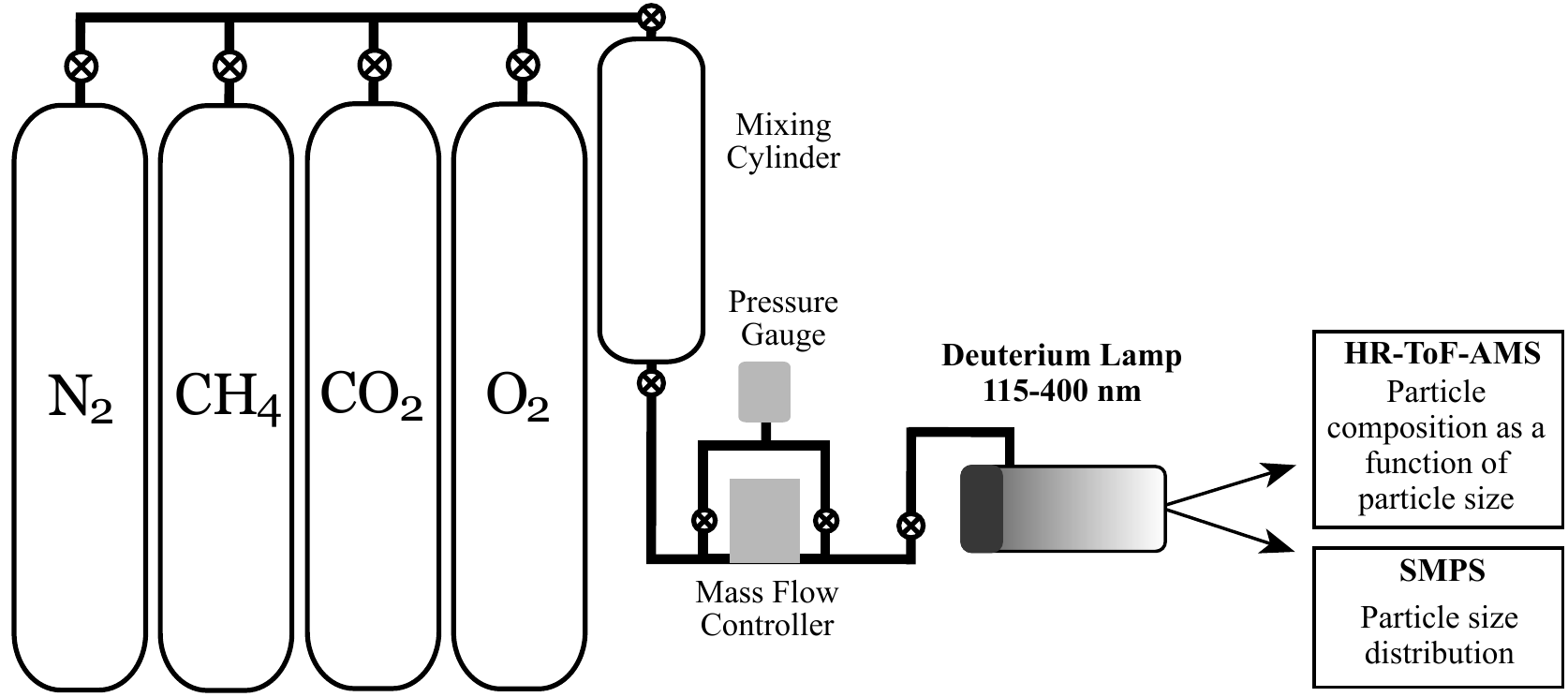}}
\caption{Shown here is a schematic of the experimental setup used for the experiments presented here. The gases mix overnight in a stainless steel mixing cylinder. Gases flow through the reaction cell where they are exposed to FUV photons from a deuterium lamp resulting in the formation of gas and solid phase products, which then flow out of the cell and into a high resolution time-of-flight aerosol mass spectrometer (HR-ToF-AMS) or a scanning mobility particle sizer (SMPS). The HR-ToF-AMS measures particle composition as a function of particle size and the SMPS measures particle size distribution. All experiments presented here used a flow rate of 100 sccm, and were run at room temperature and 620-640 torr (Boulder, Colorado atmospheric pressure). \label{fig:experiment}}
\end{figure}

A schematic of the experimental setup is shown in Figure \ref{fig:experiment}. We introduced reactant gases (CH$_{4}$ (99.99\%), CO$_{2}$ (99.99\%), O$_{2}$ (99.995\%) Airgas) into a stainless steel mixing chamber which was then filled to 600 PSI with N$_{2}$ (99.999\% Airgas). The gases mix for a minimum of 8 hrs. During the experiments, reactant gases flow continuously through a glass reaction cell with a flow rate controlled by a Mykrolis FC-2900 mass flow controller and determined by instrument requirements (100 standard cubic centimeters per minute (sccm)). The pressure in the reaction cell is maintained between 620 and 640 torr at room temperature. As they flow through the reaction cell, the gases are exposed to a flux of FUV photons from a deuterium continuum lamp (Hamamatsu L1835, MgF$_{2}$ window, 115-400 nm). The flow continues out of the reaction cell and into either a scanning mobility particle sizer (SMPS) or a high-resolution time-of-flight aerosol mass spectrometer (HR-ToF-AMS). 

\begin{figure}
\centering
\resizebox{5.5in}{!}
{\includegraphics{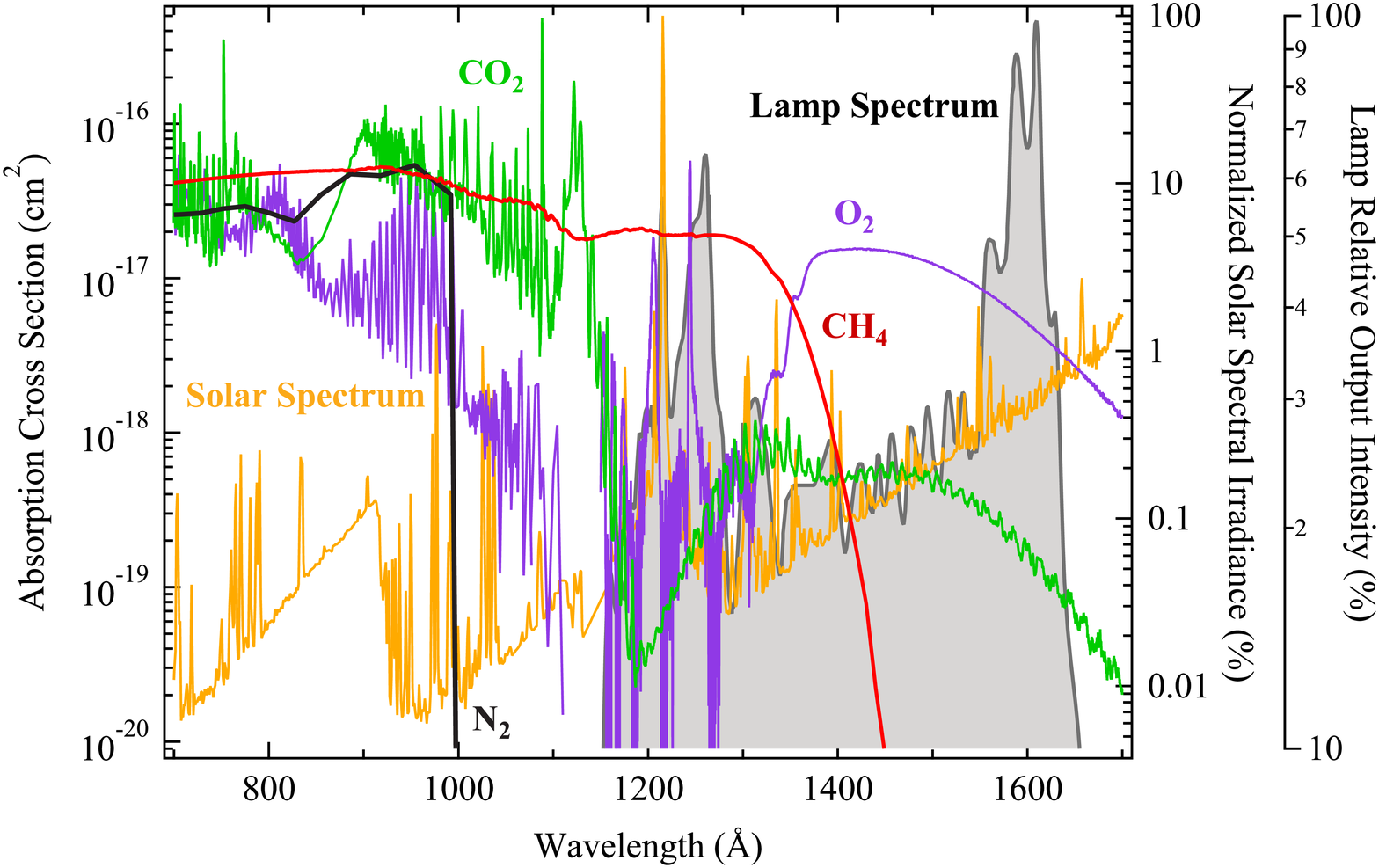}}
\caption{Shown here is a comparison of the absorption cross sections (left axis) of N$_{2}$ (black) \citep{Chan:1993d}, CH$_{4}$ (red) \citep{Chen:2004, Kameta:2002, Mount:1977}, CO$_{2}$ (green) \citep{Huestis:2010}, and O$_{2}$ (purple) \citep{Watanabe:1956, Matsunaga:1967, Lu:2010} to the solar spectrum (orange) (right, interior axis) \citep{Woods:2009} and the spectrum of the deuterium lamp used in these experiments (gray, filled) (right, exterior axis) (as provided by the manufacturer, Hamamatsu). Note that the N$_{2}$ cross section is highly structured, which is not captured in this simplified representation. There is a gap in the plotted cross sections for O$_{2}$ from 1110-1150 $\AA$; the values are much lower than the other molecules in this region.\label{fig:photo}}
\end{figure}

The FUV photons initiate chemistry that results in the formation of aerosol in addition to gas phase products. Although the photons produced by the deuterium lamp are not sufficiently energetic to directly dissociate N$_{2}$, we have demonstrated that nitrogen is participating in gas and solid phase chemistry occurring in our experiments \citep{Trainer:2012, Yoon:2014, Horst:2018}, by an as of yet undetermined mechanism that likely involves CH$_{4}$ photolysis products \citep{Trainer:2012}. Excited state atomic oxygen (e.g., O($^{1}$D)) does not react with molecular nitrogen \citep{Sander:2006} and while O$^{+}$ can dissociate N$_{2}$ \citep{Legarrec:2003}, the photons used here are not sufficiently energetic to produce O$^{+}$ directly from O$_{2}$. \citet{Prasad:2004} and \citet{Prasad:2008} find that excited states of O$_{2}$ and O$_{3}$ can react with N$_{2}$ to produce odd nitrogen species; further work is necessary to ascertain if the mechanisms they suggest are possible under our experimental conditions. Figure \ref{fig:photo} compares the lamp spectrum, solar spectrum, and cross sections of the molecules present in our initial gas mixtures.

Due to the reactive nature of the gas phase products, in particular O$_{2}$ and its photolysis products, the stainless steel tubing connecting the reaction chamber to the instruments was replaced after each experiment. The lamp's MgF$_{2}$ window was cleaned using a 60/40 mixture (by volume) of acetone and methanol between each experiment. Both of these measures were implemented to ensure that if aerosol was deposited on surfaces in the experimental apparatus during one experiment, it did not participate in the chemistry of subsequent experiments. The chamber and lines were then pumped down overnight before running the experiment. 

The instruments used to characterize the aerosol products of these experiments and data analysis techniques have been previously described by \citet{Horst:2013} and \citet{Horst:2018} (and references therein). Briefly, we use an HR-ToF-AMS (referred to hereafter as AMS; Aerodyne Research) to measure particle composition. The particles enter the AMS through a critical orifice (120 $\mu$m) and are then focused by an aerodynamic lens. The particles are flash vaporized at $\sim$600$^{\circ}$C and the resulting molecules are ionized via 70 eV electron ionization. Analysis of the ions is performed with a high-resolution time-of-flight mass spectrometer (H-TOF platform, Tofwerk). The data presented here were collected using the ÒW-modeÓ of the HR-ToF-AMS, which has a longer flight path than the ÒV-modeÓ resulting in better mass resolution (M/$\Delta$M $\sim$3000-4300 for m/z $<$200) but lower sensitivity \citet{decarlo:2006}. The AMS data were analyzed using a combination of custom software and the AMS analysis software programs SQUIRREL and PIKA \citep{decarlo:2006, Aiken:2007, Aiken:2008}. 

We use a SMPS (TSI) to measure the number of particles produced as a function of their mobility diameter. The SMPS ionizes the particles and then applies an electric field to the polydisperse aerosol and size selects the particles based on their electrical mobility against the drag force provided by a sheath flow. The size selected particles are then counted using light scattering measurements. The SMPS requires a higher flow rate than the HR-ToF-AMS so an additional flow of N$_{2}$ (99.999\% Airgas) was added after the reaction chamber to bring the total flow rate to 260 sccm. The additional flow was accounted for during data analysis.

For all gas mixtures (see Table \ref{table:exp}), identical experiments were performed using the spark discharge from a tesla coil as the energy source instead of the deuterium lamp (see e.g., \citet{Horst:2018}). However, a detectable quantity of particles was not generated. The most likely reason for the lack of particle generation is that there is not a sufficient amount of CH$_{4}$ in the gas mixture. Previous work has shown that the optimum amount of CH$_{4}$ in CH$_{4}$/N$_{2}$ gas mixtures for particle production is 2\% CH$_{4}$ in our setup using spark discharge to initiate chemistry \citep{Horst:2013}. It is therefore not surprising that these experiments did not produce a detectable quantity of particles.  

\begin{table}
\begin{center}
\caption{Summary of Experiments\label{table:exp}}
\begin{tabular}{c  | c || c c || p{0.5cm} p{0.5cm} | p{0.5cm} p{0.5cm} | p{0.5cm} p{0.5cm} | p{0.5cm} p{0.5cm}}
\hline
&CH$_{4}$& \multicolumn{2}{c||}{CH$_{4}$/CO$_{2}$}&\multicolumn{8}{c}{CH$_{4}$/CO$_{2}$/O$_{2}$}\\
\hline
$[$CH$_{4}]$& & \multicolumn{2}{c||}{$[$CO$_{2}]$}&\multicolumn{8}{c}{$[$O$_{2}]$}\\
&&260&394&\multicolumn{2}{c}{2}&\multicolumn{2}{c}{20}&\multicolumn{2}{c}{200}&\multicolumn{2}{c}{2000}\\
&&ppm&ppm&\multicolumn{2}{c}{ppm}&\multicolumn{2}{c}{ppm}&\multicolumn{2}{c}{ppm}&\multicolumn{2}{c}{ppm}\\
\hline
20 ppm&AS&\cellcolor[gray]{0.9}&\cellcolor[gray]{0.8}&\cellcolor[gray]{0.9}&\cellcolor[gray]{0.8}&\cellcolor[gray]{0.9}&\cellcolor[gray]{0.8}&\cellcolor[gray]{0.9}&\cellcolor[gray]{0.8}&\cellcolor[gray]{0.9}&\cellcolor[gray]{0.8}\\

79 ppm&AS&\cellcolor[gray]{0.9}AS&\cellcolor[gray]{0.8}AS&\cellcolor[gray]{0.9}AS&\cellcolor[gray]{0.8}AS&\cellcolor[gray]{0.9}AS&\cellcolor[gray]{0.8}AS&\cellcolor[gray]{0.9}S&\cellcolor[gray]{0.8}S&\cellcolor[gray]{0.9}S&\cellcolor[gray]{0.8}S\\

118 ppm&AS&\cellcolor[gray]{0.9}AS&\cellcolor[gray]{0.8}AS&\cellcolor[gray]{0.9}AS&\cellcolor[gray]{0.8}AS&\cellcolor[gray]{0.9}AS&\cellcolor[gray]{0.8}AS&\cellcolor[gray]{0.9}AS&\cellcolor[gray]{0.8}S&\cellcolor[gray]{0.9}S&\cellcolor[gray]{0.8}S\\

158 ppm&AS&\cellcolor[gray]{0.9}AS&\cellcolor[gray]{0.8}AS&\cellcolor[gray]{0.9}AS&\cellcolor[gray]{0.8}AS&\cellcolor[gray]{0.9}AS&\cellcolor[gray]{0.8}AS&\cellcolor[gray]{0.9}AS&\cellcolor[gray]{0.8}AS&\cellcolor[gray]{0.9}&\cellcolor[gray]{0.8}\\
\hline
\multicolumn{12}{l}{The remaining gas is N$_{2}$, bringing the total to 100\%.}\\
\multicolumn{12}{l}{S indicates measurements were made with the SMPS.}\\ 
\multicolumn{12}{l}{A indicates measurements were made with the HR-ToF-AMS.}\\
\multicolumn{12}{l}{CO$_{2}$ is indicated by column color; experiments using 260 ppm}\\
\multicolumn{12}{l}{CO$_{2}$ are light gray and experiments using 394 ppm CO$_{2}$ are}\\
\multicolumn{12}{l}{dark gray}\\
\end{tabular}
\end{center}
\end{table}

\section{Results and Discussion}

\subsection{Production rate, particle size, and particle number density} %done
\begin{figure}
\resizebox{6.1in}{!}
{\includegraphics{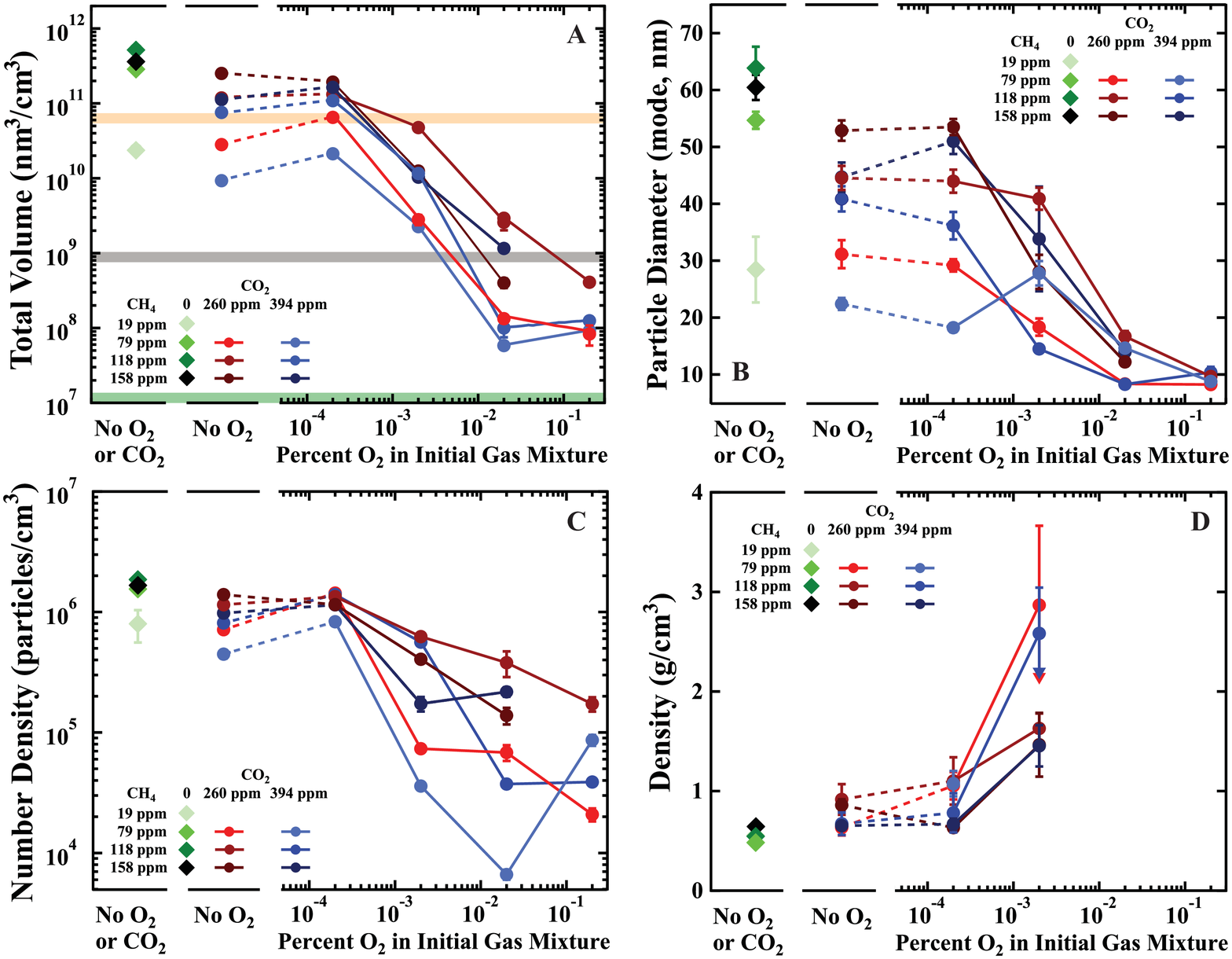}}
\caption{The total volume of particles produced for all of the experiments listed in Table \ref{table:exp} measured by the SMPS are shown in Panel A. The shaded lines indicate the production rate from our standard Titan experiment using 0.1\% CH$_{4}$ \citep{Horst:2013, Horst:2018} (orange), laboratory air during one of the experimental runs (gray), and standard background measurement (the experiment running as usual, but without the lamp on) (green). Panel B shows the particle diameter and Panel C shows the number density of particles both from SMPS measurements. Panel D presents the particle density measurements obtained by combining SMPS and AMS data. The green points are for the CH$_{4}$/N$_{2}$ experiments, while the red and blue points indicate the experiments that contained 260 and 394 ppm CO$_{2}$, respectively. }
\label{fig:SMPS_combined}
\end{figure}

Due to the complexities of multi-component gas mixtures, we first looked only at CH$_{4}$/N$_{2}$ and then at CH$_{4}$/CO$_{2}$/N$_{2}$ mixtures before adding O$_{2}$, as shown in Table \ref{table:exp}. For the CH$_{4}$/N$_{2}$ mixtures, the production rate trend is consistent with previous work using our experimental setup \citep{Trainer:2006, Horst:2013}, with an increase in production rate with increasing methane mixing ratio observed until the 158 ppm CH$_{4}$ case. The addition of CO$_{2}$ to the experiments decreases production rate and the effect is stronger for the 394 ppm case as shown in Figure \ref{fig:SMPS_combined}. Given that the highest C/O ratio in our experiments was $\sim$0.8 this effect of CO$_{2}$ is consistent with the results of \citet{Trainer:2004b} and \citet{Trainer:2006} even though these experiments use lower absolute CH$_{4}$ and CO$_{2}$ mixing ratios. At higher C/O ratios the addition of CO$_{2}$ has been shown to increase the production rate \citep{Trainer:2004b, Trainer:2006}.

\begin{deluxetable}{ll}
\caption{SMPS Measurements}
\tablehead{
\colhead{Column number} & \colhead{Column description}
}
\startdata
1 & Diameter midpoint (nm) \\
2 & 19.7 ppm CH$_{4}$ \\
3 & 78.8 ppm CH$_{4}$ \\
4 & 78.8 ppm CH$_{4}$, 260 ppm CO$_{2}$ \\
5 & 78.8 ppm CH$_{4}$, 260 ppm CO$_{2}$, 2 ppm O$_{2}$ \\
6 & 78.8 ppm CH$_{4}$, 260 ppm CO$_{2}$, 20 ppm O$_{2}$ \\
7 & 78.8 ppm CH$_{4}$, 260 ppm CO$_{2}$, 200 ppm O$_{2}$ \\
8 & 78.8 ppm CH$_{4}$, 260 ppm CO$_{2}$, 2000 ppm O$_{2}$ \\
9 & 78.8 ppm CH$_{4}$, 394 ppm CO$_{2}$ \\
10 & 78.8 ppm CH$_{4}$, 394 ppm CO$_{2}$, 2 ppm O$_{2}$ \\
11 & 78.8 ppm CH$_{4}$, 394 ppm CO$_{2}$, 20 ppm O$_{2}$ \\
12 & 78.8 ppm CH$_{4}$, 394 ppm CO$_{2}$, 200 ppm O$_{2}$ \\
13 & 78.8 ppm CH$_{4}$, 394 ppm CO$_{2}$, 2000 ppm O$_{2}$ \\
14 & 118 ppm CH$_{4}$ \\
15 & 118 ppm CH$_{4}$, 260 ppm CO$_{2}$ \\
16 & 118 ppm CH$_{4}$, 260 ppm CO$_{2}$, 2 ppm O$_{2}$ \\
17 & 118 ppm CH$_{4}$, 260 ppm CO$_{2}$, 20 ppm O$_{2}$ \\
18 & 118 ppm CH$_{4}$, 260 ppm CO$_{2}$, 200 ppm O$_{2}$ \\
19 & 118 ppm CH$_{4}$, 260 ppm CO$_{2}$, 2000 ppm O$_{2}$ \\
20 & 118 ppm CH$_{4}$, 394 ppm CO$_{2}$ \\
21 & 118 ppm CH$_{4}$, 394 ppm CO$_{2}$, 2 ppm O$_{2}$ \\
22 & 118 ppm CH$_{4}$, 394 ppm CO$_{2}$, 20 ppm O$_{2}$ \\
23 & 118 ppm CH$_{4}$, 394 ppm CO$_{2}$, 200 ppm O$_{2}$ \\
24 & 118 ppm CH$_{4}$, 394 ppm CO$_{2}$, 2000 ppm O$_{2}$ \\
25 & 158 ppm CH$_{4}$ \\
26 & 158 ppm CH$_{4}$, 260 ppm CO$_{2}$ \\
27 & 158 ppm CH$_{4}$, 260 ppm CO$_{2}$, 2 ppm O$_{2}$ \\
28 & 158 ppm CH$_{4}$, 260 ppm CO$_{2}$, 20 ppm O$_{2}$ \\
29 & 158 ppm CH$_{4}$, 260 ppm CO$_{2}$, 200 ppm O$_{2}$ \\
30 & 158 ppm CH$_{4}$, 394 ppm CO$_{2}$ \\
31 & 158 ppm CH$_{4}$, 394 ppm CO$_{2}$, 2 ppm O$_{2}$ \\
32 & 158 ppm CH$_{4}$, 394 ppm CO$_{2}$, 20 ppm O$_{2}$ \\
33 & 158 ppm CH$_{4}$, 394 ppm CO$_{2}$, 200 ppm O$_{2}$ \\
\enddata
\tablecomments{Table 2 is published in its entirety in the electronic 
edition of the {\it Astrophysical Journal}.  A portion is shown here 
for guidance regarding its form and content. The units for columns 2-33 are dN/dlogD$_{m}$ (particles/cm$^{3}$).\label{table:smps}}
\end{deluxetable}

\begin{figure}
\centering
\resizebox{5.5in}{!}
{\includegraphics{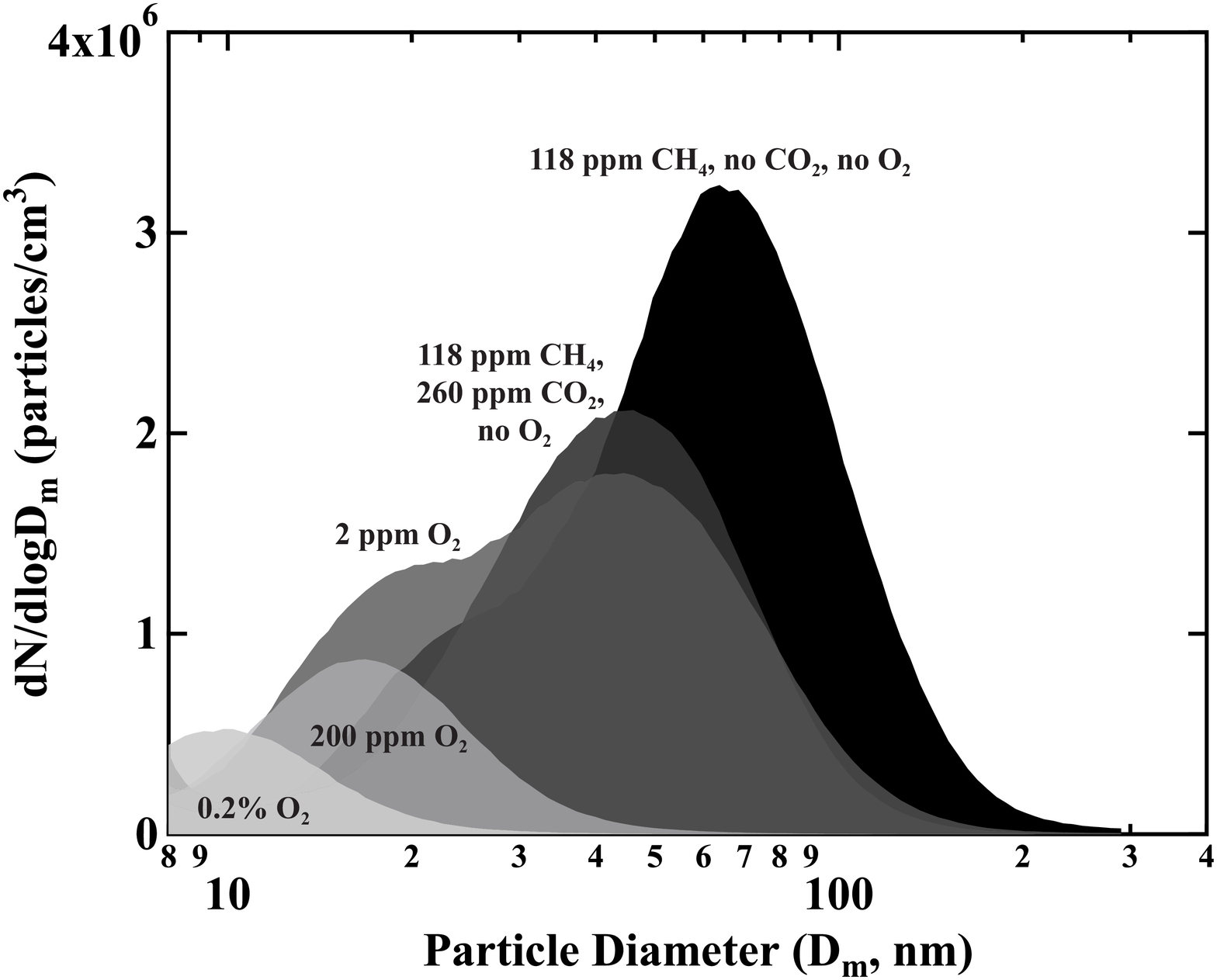}}
\caption{The size distributions for some of the 118 ppm CH$_{4}$ experiments measured by the SMPS are shown here. Of particular note is the bimodal distribution observed for the 2 ppm O$_{2}$ case. The SMPS size distribution measurements shown here and for all other experiments with sufficiently high production rates are available in Table \ref{table:smps}.\label{fig:smps_o2}}
\end{figure}

For all but the 260 ppm CO$_{2}$/158 ppm CH$_{4}$ case, the addition of 2 ppm O$_{2}$ increased the production rate, as shown by the increasing particle volume in the SMPS (Panel A of Figure \ref{fig:SMPS_combined}). However, the addition of more O$_{2}$ (20 ppm or greater) resulted in a decrease in aerosol production rate with increasing O$_{2}$ concentration. As demonstrated by the SMPS measurements, the observed trends in aerosol loading shown in Panel A of Figure \ref{fig:SMPS_combined} result from a decrease in particle diameter (Panel B) and a decrease in the number of particles (Panel C) with increasing O$_{2}$ concentration, indicating both nucleation and growth are affected by the inclusion of O$_{2}$ in the gas mixtures. As shown in Figures \ref{fig:SMPS_combined} (Panel B) and \ref{fig:smps_o2}, as the O$_{2}$ concentration increases the particle diameter decreases, which has a few consequences for our measurements. First, it means that for our high oxygen concentration cases, we are likely missing part of the aerosol distribution at the small diameter end (e.g., the 0.2\% O$_{2}$ size distribution in Figure \ref{fig:smps_o2}) so the loading measurements shown in Figure \ref{fig:SMPS_combined} represent a lower limit. This will not be a large effect as it is the small radius end of the distribution that is missing. Second, there are implications for the density measurements, discussed below.

Assuming the particles maintain the same optical properties, the transition to smaller particle size with increasing O$_{2}$ pushes the particles further into the Rayleigh scattering regime, where they will scatter shorter wavelengths more efficiently. Given the changes in composition discussed below, it is likely that the particles change both size and optical properties indicating that the effect on radiative transfer in the atmosphere will require measurement of optical constants and additional modeling to understand.

We would also like to note that we did not include H$_{2}$O in the initial gas mixture so that we could focus on the previously unexplored effect of O$_{2}$, however models indicate that there may have been water at the 100 ppm level even in the stratosphere \citep{Pavlov:2000}. To our knowledge, the effect of gas phase H$_{2}$O in the absence of liquid water has not been experimentally explored previously with the exception of \citet{Horst:2018exo}, \citet{He:2018}, and \citet{He:2018b}, which found that gas mixtures with substantial amounts of H$_{2}$O (55-66\%) were favorable for haze formation. 

The effect of oxygen bearing species on photochemical haze generation is complicated by competing processes. While the addition of oxygen bearing functional groups will terminate hydrocarbon chains (therefore often slowing or terminating growth) decreasing the solid production rate, the presence of oxygen bearing functional groups also tends to lower the vapor pressure of species resulting in more efficient partitioning into the solid phase (see excellent discussion in \citet{Trainer:2006}). Further, H$_{2}$ has been shown to decrease haze production rates \citep{Raulin:1982, dewitt:2009, Sciamma:2010} and the presence of oxygen bearing species may serve to remove some H$_{2}$ from the system reducing that effect (see e.g., \citet{Trainer:2004b, Trainer:2006, Horst:2014, He:2017, Horst:2018exo}). 

Differences in the absorption cross sections of oxygen species, as shown for O$_{2}$ and CO$_{2}$ in Figure \ref{fig:photo}, further complicates identification of the mechanisms responsible for our observed changes. Both O$_{2}$ and CO$_{2}$ have larger absorption cross sections than CH$_{4}$ at longer wavelengths and their presence could reduce photons that would otherwise be available to photolyze larger hydrocarbons and nitriles resulting in the production of larger molecules. At shorter wavelengths where CH$_{4}$ absorbs, CO$_{2}$ and O$_{2}$ also have cross sections that are competitive with CH$_{4}$ and thus their presence also likely decreases the efficiency of CH$_{4}$ photolysis. The decrease in CH$_{4}$ photolysis represents a decrease in the very first step of haze formation. These experiments were not designed to disentangle these different effects, but it is important to keep them in mind.

\subsection{Particle density} %done
Particle density is an important parameter in cloud microphysics, but is often set to one when modeling planetary hazes for simplicity and lack of any other constraints. Previous work has shown that this assumption may substantially overestimate particle density for Titan \citep{Horst:2013}. Figure \ref{fig:SMPS_combined} (Panel D) shows particle density, which is calculated using a combination of size measurements from the SMPS ($D_{m}$) and particle time-of-flight (PToF) mode of the AMS ($D_{va}$) as discussed extensively in \citet{DeCarlo:2004} and \citet{Horst:2013}. Previous works have shown that the two diameters are related by effective particle density $\rho_{eff}$,
\begin{equation} \rho_{eff}=\rho_{0}\frac{D_{va}}{D_{m}}=\rho_{m}S \end{equation}
where $\rho_{0}$ is the unit density (1 g/cm$^{3}$), $\rho_{m}$ is the material density of the particle, and $S$ is the shape factor \citep{DeCarlo:2004, Jimenez:2003a, Jimenez:2003b}.

At the small particle end of the size distribution, the SMPS and AMS differ in their ability to detect particles, with the SMPS being much more effective. The aerodynamic lens in the AMS transmits particles larger than 20 nm, but the greatest efficiency is for particles larger than 60 nm in diameter \citep{Jayne:2000}. For this reason, Figure \ref{fig:SMPS_combined} does not have density measurements for the 200 ppm and 0.2\% O$_{2}$ cases, even though particles were measured, because their size distributions were not reliably measured by the AMS. This is also responsible for the larger error bars for the 20 ppm O$_{2}$ cases. Note that for the 79 ppm CH$_{4}$, 20 ppm O$_{2}$ experiments, due to the AMS transmission efficiency, $D_{va}$ is an upper limit and therefore the densities indicated are also upper limits. In general, particle densities increased with the addition of CO$_{2}$ as compared to the particles made only from N$_{2}$/CH$_{4}$ in the initial gas mixtures, and the densities further increased with the addition of O$_{2}$ to the experiments. This is consistent with the behavior observed in \citet{He:2017} where density increased with increasing CO in the initial gas mixture indicating that particles with a higher weight percent of oxygen are more dense. \citet{He:2017} note that their density increase was too large to be explained solely by the large molecular weight of the atomic oxygen compared to carbon, nitrogen, and hydrogen and could be related to other factors such as polarity resulting in increased intermolecular forces. 

\subsection{Particle composition} 

\begin{figure}
\centering
\resizebox{5.5in}{!}
{\includegraphics{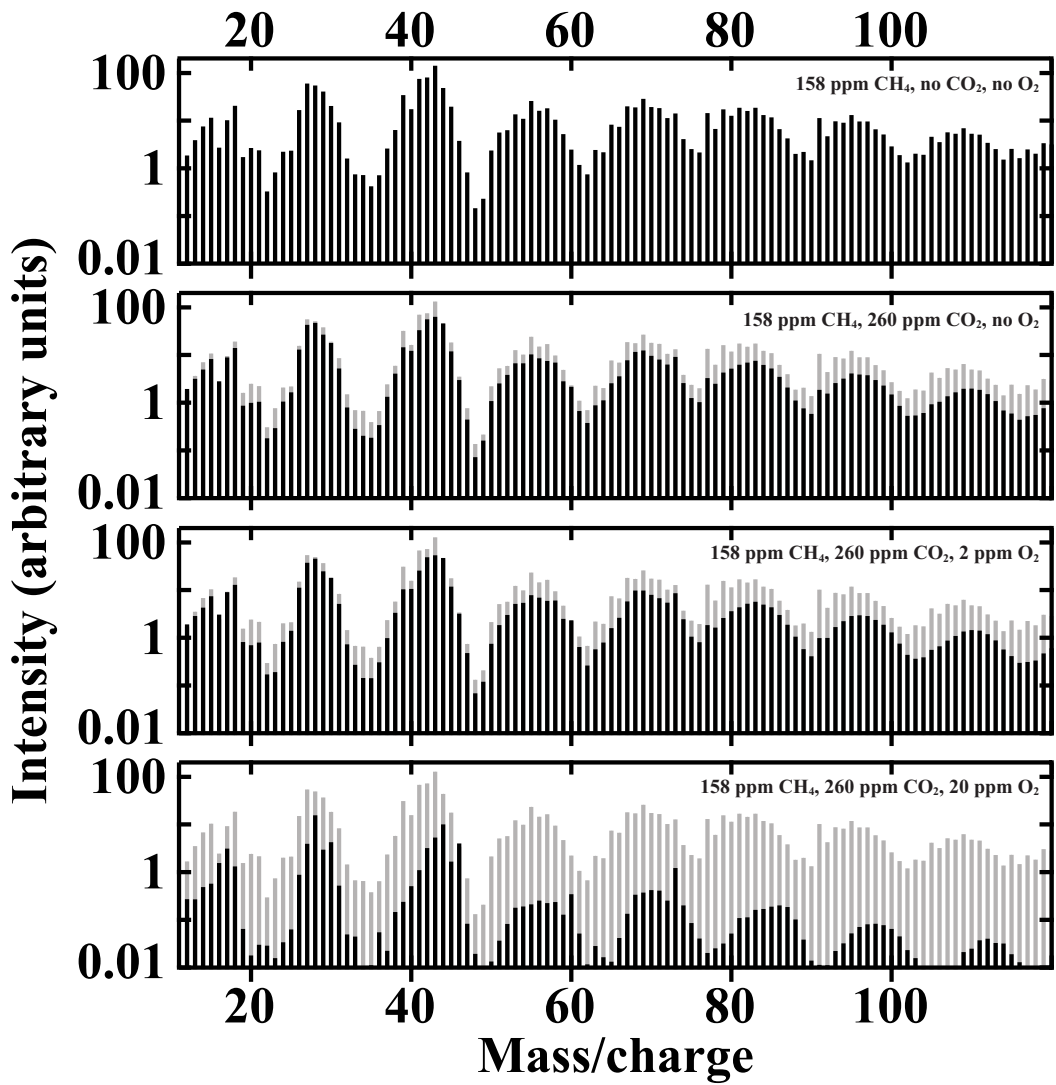}}
\caption{Here we show how the HR-ToF-AMS measurements change with the addition of CO$_{2}$ and O$_{2}$ using the 158 ppm CH$_{4}$ and 260 ppm CO$_{2}$ cases. To facilitate comparison, the 158 ppm CH$_{4}$ measurements (top) are shown in gray with the rest of the measurements. The processed HR-ToF-AMS measurements shown here and for all other experiments with sufficiently high production rates are available in Table \ref{table:ams}. \label{fig:ms}}
\end{figure}

Figure \ref{fig:ms} shows the AMS measurements for the 158 ppm CH$_{4}$/260 ppm CO$_{2}$ experiments (shown at unit mass resolution for simplicity). The addition of CO$_{2}$ causes an obvious decrease in the relative amounts of benzene (78) and toluene (91). This decrease in aromatic molecule production is consistent with that observed by \citet{Trainer:2006}, which used larger concentrations of CH$_{4}$ and CO$_{2}$. This observation is particularly important given that aromatic molecules are efficient absorbers in the UV and visible, and experiments have shown that their presence impacts the absorption of haze analog particles in the far-infrared \citet{Trainer:2013, Sebree:2014, Sciamma:2017}. Therefore the decrease in aromaticity may impact the optical properties.  

\begin{deluxetable}{ll}
\caption{HR-ToF-AMS Measurements}
\tablehead{
\colhead{Column number} & \colhead{Column description}
}
\startdata
1 & Mass/charge \\
2 & 19.7 ppm CH$_{4}$ \\
3 & 78.8 ppm CH$_{4}$ \\
4 & 78.8 ppm CH$_{4}$, 260 ppm CO$_{2}$ \\
5 & 78.8 ppm CH$_{4}$, 260 ppm CO$_{2}$, 2 ppm O$_{2}$ \\
6 & 78.8 ppm CH$_{4}$, 260 ppm CO$_{2}$, 20 ppm O$_{2}$ \\
7 & 78.8 ppm CH$_{4}$, 394 ppm CO$_{2}$ \\
8 & 78.8 ppm CH$_{4}$, 394 ppm CO$_{2}$, 2 ppm O$_{2}$ \\
9 & 78.8 ppm CH$_{4}$, 394 ppm CO$_{2}$, 20 ppm O$_{2}$ \\
10 & 118 ppm CH$_{4}$ \\
11 & 118 ppm CH$_{4}$, 260 ppm CO$_{2}$ \\
12 & 118 ppm CH$_{4}$, 260 ppm CO$_{2}$, 2 ppm O$_{2}$ \\
13 & 118 ppm CH$_{4}$, 260 ppm CO$_{2}$, 20 ppm O$_{2}$ \\
14 & 118 ppm CH$_{4}$, 394 ppm CO$_{2}$ \\
15 & 118 ppm CH$_{4}$, 394 ppm CO$_{2}$, 2 ppm O$_{2}$ \\
16 & 118 ppm CH$_{4}$, 394 ppm CO$_{2}$, 20 ppm O$_{2}$ \\
17 & 158 ppm CH$_{4}$ \\
18 & 158 ppm CH$_{4}$, 260 ppm CO$_{2}$ \\
19 & 158 ppm CH$_{4}$, 260 ppm CO$_{2}$, 2 ppm O$_{2}$ \\
20 & 158 ppm CH$_{4}$, 260 ppm CO$_{2}$, 20 ppm O$_{2}$ \\
21 & 158 ppm CH$_{4}$, 394 ppm CO$_{2}$ \\
22 & 158 ppm CH$_{4}$, 394 ppm CO$_{2}$, 2 ppm O$_{2}$ \\
23 & 158 ppm CH$_{4}$, 394 ppm CO$_{2}$, 20 ppm O$_{2}$ \\
\enddata
\tablecomments{Table 3 is published in its entirety in the electronic 
edition of the {\it Astrophysical Journal}.  A portion is shown here 
for guidance regarding its form and content. Column 1 is mass/charge. All other columns are intensity in arbitrary (but internally consistent) units.\label{table:ams}}
\end{deluxetable}

\begin{figure}
\centering
\resizebox{5.5in}{!}
{\includegraphics{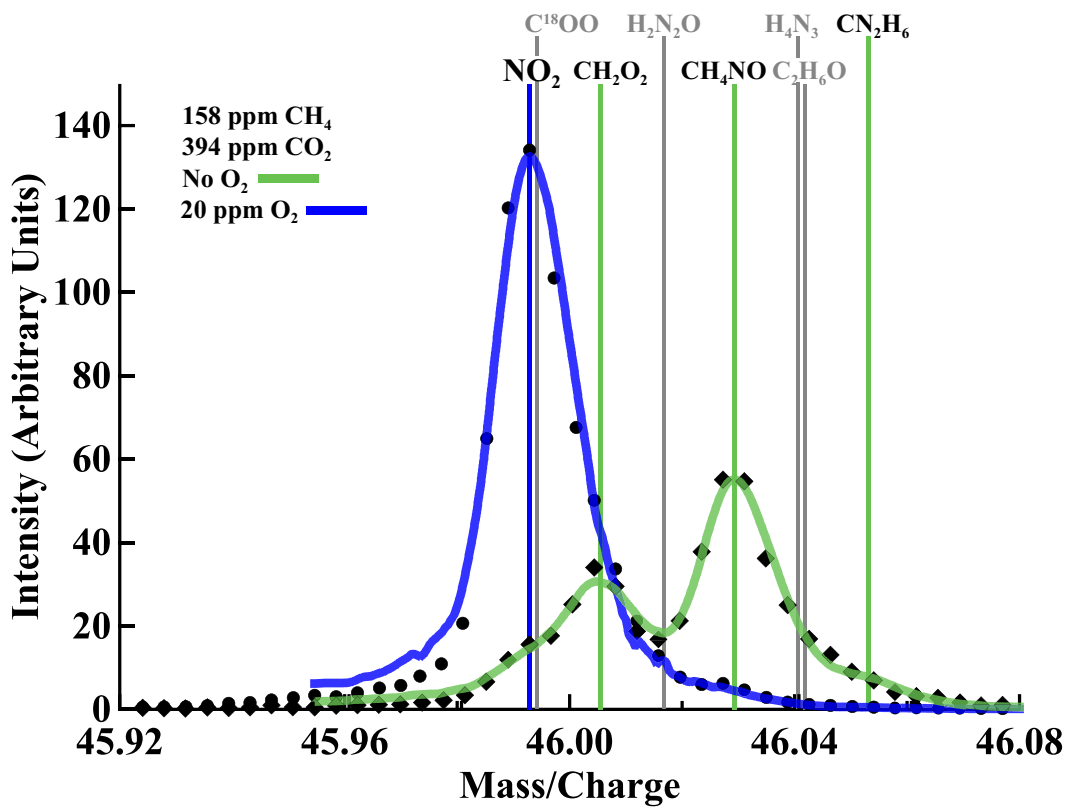}}
\caption{The HR-ToF-AMS has sufficient resolution and accuracy to differentiate between NO$_{2}$ and other molecules with the same nominal mass (46) as shown here. The thick green line shows the total fit to the data for the 158 ppm CH$_{4}$, 394 ppm CO$_{2}$ experiment, while the thick blue line shows the total fit when 20 ppm O$_{2}$ is added. The possible CHNO containing peaks with a nominal mass of 46 are also shown; those in gray were not used to fit the data from either experiment. The fit to the 20 ppm O$_{2}$ case includes only NO$_{2}$ (thick blue line). The fit to the no O$_{2}$ experiment includes NO$_{2}$, plus CH$_{2}$O$_{2}$, CH$_{4}$NO, and CN$_{2}$H$_{2}$. Note that although C$^{18}$OO has almost the identical mass as NO$_{2}$, its abundance is constrained by the parent peak. Although the intensity here is in arbitrary units, the comparison between the two experiments is quantitatively valid. \label{fig:no2}}
\end{figure}

\begin{figure}
\centering
\resizebox{5.5in}{!}
{\includegraphics{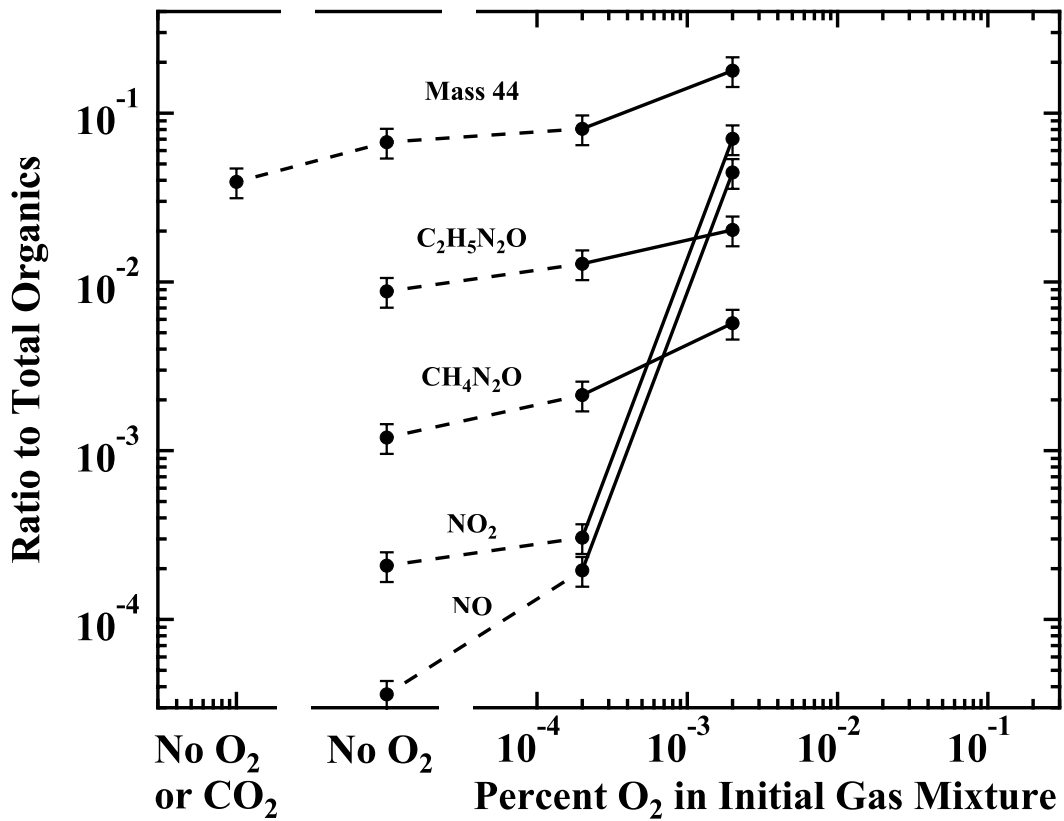}}
\caption{Shown here are the ratios of the mass 44 peak, NO, NO$_{2}$, CH$_{4}$N$_{2}$O, and C$_{2}$H$_{5}$N$_{2}$O to the total organic content of the aerosol from AMS measurements for the 158 ppm CH$_{4}$ measurements shown in Figure \ref{fig:ms}. \label{fig:mass44}}
\end{figure}

The addition of CO$_{2}$ and then O$_{2}$ results in the formation of newly prominent peaks, which are most obvious in Figure \ref{fig:ms} at mass 46, 60, and 73. The mass resolution of the HR-ToF-AMS is sufficient to allow for identification of the specific mass responsible for the changes measured at these peaks and the changes are attributed to increases in NO$_{2}$ (46) (see Figure \ref{fig:no2}), CH$_{4}$N$_{2}$O (60), and C$_{2}$H$_{5}$N$_{2}$O (73), respectively. It is interesting to note that CH$_{4}$N$_{2}$O is the molecular formula for urea. NO (at mass 30) also increases with increasing O$_{2}$, although it does not stand out as strongly. The ratio of these peaks to the total organic signal is shown in Figure \ref{fig:mass44}. The NO and NO$_{2}$ peaks likely result from fragmentation of nitrate bearing molecules by the electron ionization system in the AMS (see e.g., \citet{Farmer:2010}). It can be challenging to differentiate between organic and inorganic nitrate in the AMS and none of the methods explored by \citet{Farmer:2010} appear to be appropriate for our data, presumably due to the very different types of aerosol investigated. The presence of numerous C$_{x}$H$_{y}$N$_{z}$O$_{p}^{+}$ ions indicates that some, if not all, of the observed nitrate is organic. Regardless of the final form, it is clear that the haze formation processes observed in these experiments are converting atmospheric N$_{2}$ to ``fixed" nitrogen species, which are required for life and may have been the limiting nutrient during points in Earth's history (see discussion in \citet{Trainer:2013earth}). Atmospheric production of fixed nitrogen is particularly interesting because it provides a global source, including a source of fixed nitrogen for Earth's oceans, as discussed previously by \citet{Yung:1979} and \citet{Tian:2011}. 

\begin{figure}
\centering
\resizebox{5.5in}{!}
{\includegraphics{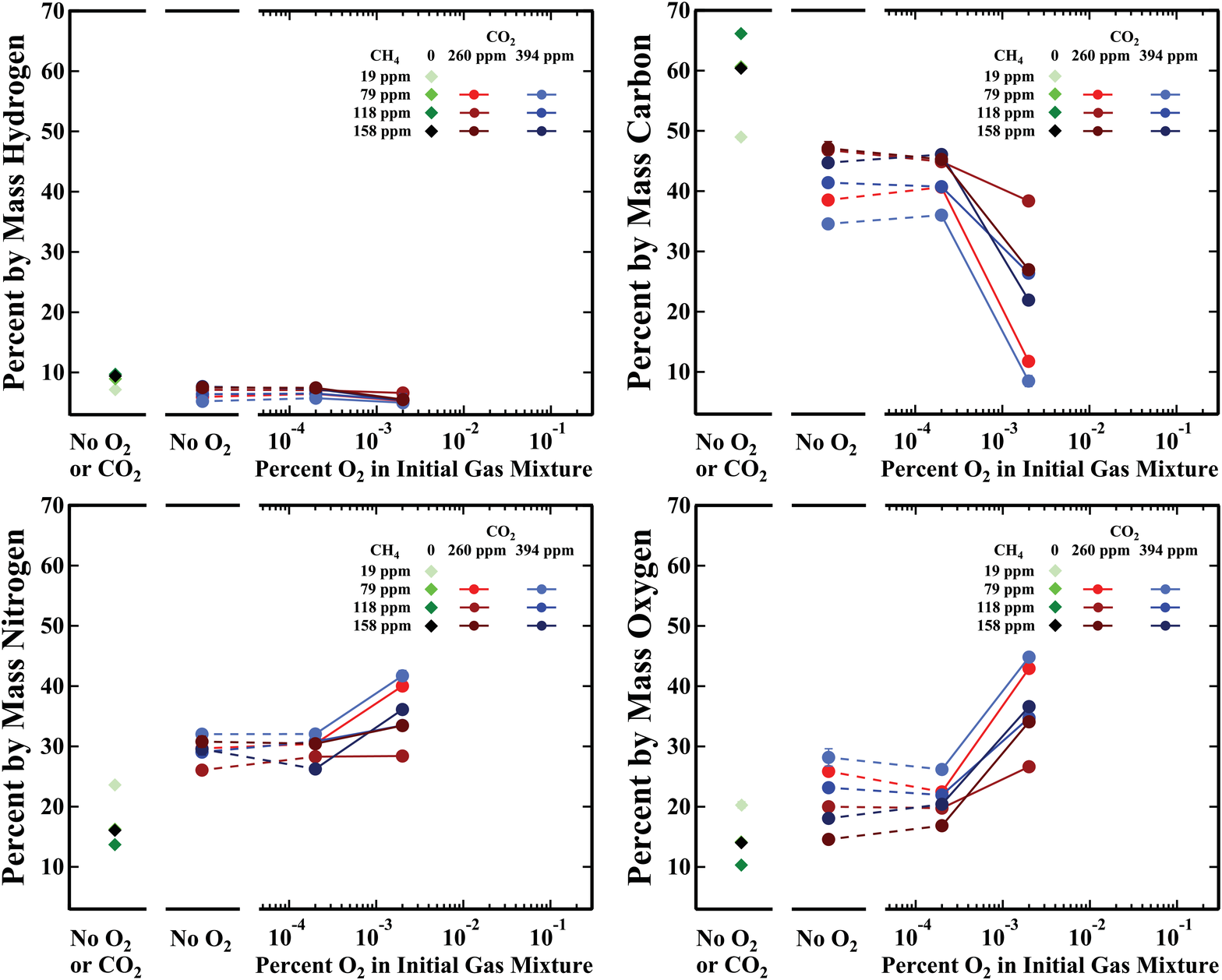}}
\caption{Shown here is the percent by mass of hydrogen, carbon, nitrogen, and oxygen in the aerosol particles as measured by the HR-ToF-AMS. Error bars are 1 sigma on the reproducibility of the measurements and are generally smaller than the symbol size. See text for further error discussion. \label{fig:nitrogen}}
\end{figure}

Previous work (see e.g., \citet{Trainer:2004b}) has used the peak at mass 44 as a proxy for degree of oxidation of organics in AMS measurements because of the presence of COO$^{+}$, which is a typical fragment of oxidized organic molecules. Indeed Figure \ref{fig:mass44} shows that the addition of CO$_{2}$, and then the addition of O$_{2}$, results in an increase of the ratio of mass 44 to the total organic signal indicating that the aerosol becomes more oxidized. The increase in oxidation has implications both for the classes of compounds delivered to the surface and the optical properties of the haze particles. The trend observed with mass 44 is consistent with the overall elemental composition calculated from the AMS measurements. As shown in Figure \ref{fig:nitrogen}, the addition of CO$_{2}$ and subsequent addition of O$_{2}$ results in an increase in the nitrogen content of the aerosol. This is true also of the oxygen content and it is at the expense of carbon. The elemental composition is also consistent with the increased abundance of nitrogen and oxygen containing compounds discussed above. Note that the error bars indicated in Figure \ref{fig:nitrogen} represent the error only on the reproducibility of the measurements. Extensive calibration of the HR-ToF-AMS with Earth atmosphere relevant compounds \citep{Aiken:2007, Aiken:2008} finds approximately 20\% error on elemental ratios but the numbers were not reported for elemental percentages and are therefore not shown here.

The increased oxidation of laboratory haze analogues produced with CO$_{2}$ in addition to CH$_{4}$ affects their optical properties, as shown by previous works. \citet{Hasenkopf:2010} found a change in extinction at 532 nm when introducing CO$_{2}$ into CH$_{4}$/N$_{2}$ system resulting from an increase in the imaginary index of refraction. \citet{Gavilan:2017} find similar results for analogues generated with a cold plasma in the UV wavelength region. However, as CO$_{2}$ is introduced in a 1:1 ratio with CH$_{4}$, the imaginary index of refraction actually decreases in the visible wavelength region that \citet{Gavilan:2017} were able to model, which is opposite of what is observed by \citet{Hasenkopf:2010}. Clearly more work is necessary to understand the effect of oxygen bearing species on optical properties of haze particles. 

\section{Summary}

Here we investigated the effect of O$_{2}$ on haze formation and composition in laboratory atmosphere simulation experiments. For the smallest amount of O$_{2}$ (2 ppm) used, the addition resulted in an increase in particle production for all gas mixtures except one. However, at higher oxygen concentrations (20 ppm and greater) the addition of O$_{2}$ decreases the production rate of haze. As the amount of O$_{2}$ present in the experiments increases, both the particle size and number of particles decreases, which indicates that O$_{2}$ affects particle nucleation and growth. Additionally, the density of the particles increases with increasing O$_{2}$. O$_{2}$ also affects the chemistry occurring in the chamber, resulting in measurable composition changes in the particle phase.

In terms of composition, the addition of the two oxygen bearing species (CO$_{2}$ and O$_{2}$) not only increased the amount of oxygen in the aerosol, but it also increased the degree of nitrogen incorporation in the aerosol. In particular, the addition of O$_{2}$ resulted in the formation of nitrate bearing molecules and other molecules that contain both nitrogen and oxygen. The fact that the presence of oxygen bearing molecules increased the efficiency of nitrogen fixation has implications for the role of haze as a source of molecules required for the origin and evolution of life. In addition to the production of molecules that are potentially biologically interesting, the observed increase in oxidation state of the aerosol has a number of consequences for the transport of radiation in the atmosphere. For example, it is likely that the more oxidized aerosol is more hygroscopic and therefore a better CCN (see e.g., \citet{Trainer:2013earth}), potentially affecting the atmospheric albedo due to changes in clouds. The composition changes also likely affect the absorption and scattering behavior of these particles but measurements of the optical properties are required to fully understand the implications for the effect on radiative transfer.  

Although the production rate decreases, the presence of modest amounts of O$_{2}$ does not fully suppress haze formation and therefore its possible existence (including effects on climate and as a source of material to the surface) should not be neglected. We note that the sensitivity of the particle production rate to the abundances of the gases investigated here indicates the possibility that the existence of a haze layer and the properties of the haze particles present may have varied over time as the absolute and relative abundances of CO$_{2}$, CH$_{4}$, and O$_{2}$ varied. 

\acknowledgments
SMH was supported in part by NSF Astronomy and Astrophysics Postdoctoral Fellowship AST-1102827. CH was supported by the Morton K. and Jane Blaustein Foundation. MSU was supported by NASA Earth and Space Sciences Fellowship NNX14AO32H.

 \bibliography{Titanoxygen}

%\begin{thebibliography}{}
%\end{thebibliography}



\end{document}